# Device identification using optimized digital footprints

**Rajarshi Roy Chowdhury, Azam Che Idris, Pg Emeroylariffion Abas**
Faculty of Integrated Technologies, Universiti Brunei Darussalam, Jalan Tungku Link, Bandar Seri Begawan, Brunei Darussalam

**ABSTRACT**

The rapidly increasing number of internet of things (IoT) and non-IoT devices has imposed new security challenges to network administrators. Accurate device identification in the increasingly complex network structures is necessary. In this paper, a device fingerprinting (DFP) method has been proposed for device identification, based on digital footprints, which devices use for communication over a network. A subset of nine features have been selected from the network and transport layers of a single transmission control protocol/internet protocol packet based on attribute evaluators in Weka, to generate device-specific signatures. The method has been evaluated on two online datasets, and an experimental dataset, using different supervised machine learning (ML) algorithms. Results have shown that the method is able to distinguish device type with up to 100% precision using the random forest (RF) classifier, and classify individual devices with up to 95.7% precision. These results demonstrate the applicability of the proposed DFP method for device identification, in order to provide a more secure and robust network.

*Corresponding Author:*

Rajarshi Roy Chowdhury
Faculty of Integrated Technologies, Universiti Brunei Darussalam
Jalan Tungku Link, BE1410, Bandar Seri Begawan, Brunei Darussalam
Email: 19h0901@ubd.edu.bn or rajarshiry@gmail.com

## 1. INTRODUCTION

The rapid technological revolution has brought plenty of new opportunities to the doorstep of innovators and entrepreneurs [1]. This is possible due to the presence of a large number of heterogeneous internet of things (IoT) and non-IoT devices connected to the Internet, which are capable of providing various services over networks [2]. However, IoT devices are commonly resource-constraint, in terms of memory, processing power and energy [3], [4], as they are built for specific-purposes. On the other hand, non-IoT devices are general-purpose computing devices [5], [6] used in everyday life for various purposes, and do not have these constraints. IHS-Markit [7] has predicted that IoT devices connected to the internet will reach approximately 125 billion by the year 2030. Proliferated growth of IoT and non-IoT devices has imposed new challenges; on network administrators and operators, to securely manage and control network services [8], [9]. Though essential for communication, using traditional explicit identifiers, such as internet protocol (IP) and media access control (MAC) addresses, to determine identities of devices that are connected to a network, is not secure, as these addresses are easily mutable. Xu *et al.* [10] have shown that commonly available software with a basic level of knowledge on networking, can be used to easily change explicit identifiers for malicious intent. As such, implicit device identifiers, such as device fingerprinting (DFP) based on network traffic, MAC frame, and radio signal, are required to address some of these security issues [10]–[12].

In this paper, an efficient DFP method has been proposed based on the analysis of network packet header features from the network and transport layers of the open systems interconnection model [13]: IP, transmission control protocol (TCP), and user datagram protocol (UDP). The proposed DFP scheme is solely based on the device-originated communication footprints, which can be captured using simple network packet analyzers. A set of 9 features have been extracted from device-originated packets header protocols, to construct fingerprints for both IoT and non-IoT devices in the proposed model for the task. This subset of features has been selected based on statistical and empirical assessments of 82 features from the TCP, UDP, and IP protocols. The method has been evaluated using different supervised machine learning (ML) classification algorithms: J48, random forest (RF), random tree (RT), naive bayes (NB), bagging (BG), and vote (VT), on publicly available datasets [5], [14]: IoT Sentinel and University of New South Wales (UNSW), as well as on an experimental testbed dataset consisting of 7 non-IoT devices. A machine learning (ML) based DFP model assists in identifying devices using only their communication traffic characteristics, which helps network administrators to identify devices even when devices randomly change their IP/MAC addresses. A supervised DFP model is required to train with a set of label instances (training dataset) and then evaluate using a training dataset (unseen instances). For preparing these training and testing datasets, device-originated traffic traces are filtered according to individual device MAC addresses and labeled with devices' name. Results have shown that the proposed DFP model achieves up to 99.0% precision in identifying individual non-IoT devices, and up to 95.7% precision in identifying both IoT and non-IoT devices, using the RF classifier. The model has also been shown to be able to distinguish device type with up to 100% precision. This would undoubtedly assist network administrators and operators in the management of a network. However, if a new device is joined in a network, the proposed DFP model is required to train with new instances from this device.

## 2. RELATED WORK

Due to its perceived potential, researchers [5], [6], [11], [14], [15] have been analyzing communication traffic traces of network-connected devices, to extract device-specific fingerprints, which may be used for classification and distinguishing between IoT and non-IoT devices in a network. A set of features can be extracted from the different layers of the communication models [13], through statistical measurements of the packets, either from the packet's payload or protocols information. Sivanathan *et al*. [16] presented a DFP approach based on statistical analysis of network traffic. Passively observed traffic traces were measured, including device sleeping time, packet size, and domain name system (DNS) intervals, to generate fingerprints. An ML algorithm was then utilized for classification. The scheme has been shown to be able to, not only distinguish device type, but also classify individual IoT devices with high accuracy. Similarly, in [5], Sivanathan *et al*. utilized eight statistical features, including flow volume, duration, rate, and device sleep time, from communication flows to identify both IoT and non-IoT devices. The model gains over 99% accuracy in identifying individual devices.

Bremler-Barr *et al*. [6] described a device identification framework, to characterize IoT and non-IoT devices using statistical attributes of network traffic from different layers of the open systems intercommunication (OSI) model. 27 features, including TCP window size, number of unique port numbers, and TCP/UDP packets ratio, were extracted from different packets header protocols to be used as DFP for classification, with each feature statistically evaluated to choose the most informative features subset. These features were then used for training an ML model. It has been shown that 100% precision is achievable by using a unified classifier, however, a long 20 minutes classification latency is required for the classification. Pinheiro *et al*. [17] presented a DFP technique for the classifications of IoT and non-IoT devices in a smart environment based on statistical features of the communication packets length. Packets were grouped into a 1 second window according to the packets timestamp. Statistical attributes, including mean and standard deviation of a group of $n$ packets length, and total number of bytes transmitted by every device in a chosen time scale, were computed to generate DFP. These attributes were then utilized for training a binary classifier for distinguishing device type, whilst a multi-class classifier was used to determine IoT devices connected in a network. The scheme achieved 99% precision in discriminating between device type, with individual IoT device classification reaching up to 96%.

Ortiz *et al*. [18] introduced a deep learning (DL) technique for the classifications of IoT and non-IoT devices using traffic traces. A DL technique using stacked auto-encoders was utilized to automatically learn a set of significant features from $n$ number of TCP-flow samples from every device. The scheme achieved good performance in identifying previously seen devices, with over 70% accuracy attained in distinguishing between device type on an unseen dataset. Meidan *et al*. [19] proposed a DFP technique, which is capable of identifying IoT and non-IoT devices using network traffic characteristics. Individual IoT device statistical features, including incoming and outgoing traffic ratio, were calculated from TCP sessions of the devices, to generate fingerprints. A multi-stage meta classifier was then trained using the feature vectors for classification. The model achieved over 99% accuracy for individual device classification using a dataset of 9 IoT devices. Furthermore, the classifier was able to distinguish device type with high accuracy. On the other hand, in [11], [20], the researchers presented DFP models for classifying devices using only packet information, whilst 82% and 83.35% accuracies gain with 212 and 161 features, respectively, including source packet length, checksum, and protocol number.

## 3. PROPOSED METHODOLOGY
### 3.1. Datasets and data collection

Two publicly available datasets [5], [14]: IoT Sentinel and UNSW datasets, and an experimental dataset, have been used in this paper as shown in Table 1. The UNSW dataset [5] consists of network traffic traces from 21 IoT (e.g. AmazoneEcho, BelkinMotion, and WithingsScale devices) and 7 non-IoT (e.g. Laptop, and SamsungTab) devices, with their communication footprints (raw packet information) captured by emulating a smart living environment for a long duration of time. On the other hand, the IoT Sentinel dataset [14] includes 31 IoT devices, including Aria, D-LinkCam, Lightfy, and iKettle2 devices, with only traffic traces during the setup phase available. Additionally, a non-IoT dataset has been collected from an experimental testbed in a laboratory environment, to verify the performance of the proposed DFP model. This dataset has been collected to supplement the relatively small number of non-IoT devices in the publicly available datasets.

Table 1. List of IoT and non-IoT datasets

| Dataset | Devices | IoT | Non-IoT | Year | Source |
|---|---|---|---|---|---|
| IoT-Sentinel | 31$^{IoT}$ | 102,240 | - | 2017 | Miettinen et al. [14] |
| UNSW | 22$^{IoT}$+7$^{Non-IoT}$ | 6,844,821 | 3,515,722 | 2017 | Sivanathan et al. [5] |
| Lab | 7$^{Non-IoT}$ | - | 442,970 | 2021 | - |

The lab dataset consists of communication traffic traces from 7 non-IoT devices of different types, including laptops, desktops, and smartphones, from different manufacturers. Figure 1 depicts the laboratory experimental setup, with the devices configured to associate with an access point. A laptop has been used to provide internet services to the devices, as well as to capture network traffic from the devices. It connects to the local network using the built-in Ethernet interface, which provides a connection to the Internet through a gateway. An Ethernet-adapter connects with the laptop, to allow a local area network (LAN) to be setup, for providing network services to the devices using a hub. The external Ethernet-interface has been configured by utilizing a network connection editor tool, to allow Ethernet setup and IPv4-Setting tabs for connection establishment. Additionally, a Hotspot has also been configured, to setup a wireless LAN (WLAN) for providing network services to WiFi-enabled devices. A shell-script has been used to automate data collection and storing processes, with tcpdump used to capture and store traffic traces passing over both the external Ethernet-interface as well as the built-in WiFi-interface, into .pcap files format. TShark has been used to extract device-originated packets features from the capture pcap files, which were then stored in .csv files format, with the names of the individual devices used as labels. The dataset has also been cleaned by removing inconsistencies: empty rows, duplicate values, whilst corrupted pcap files have been repaired by utilizing pcapfix [21].

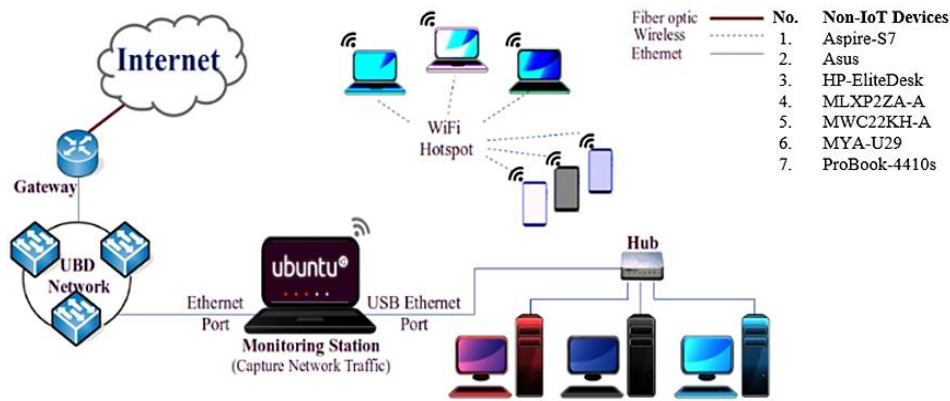

Figure 1. An experimental testbed design of a non-IoT devices network for data collection process

### 3.2. The proposed device fingerprinting method

For the purpose of this paper, captured network traffic traces may come from any of the three datasets considered or combinations thereof. Device-originated digital footprints were filtered according to MAC addresses of the devices. A list of different types of packets, including the TCP, UDP, IP, and internet control message protocol (ICMP), were utilized to extract features set for generating fingerprints. Initially, 82 features were extracted from the network and transport layers protocols header for DFP. Subsequently, all these features were assessed using an attributed evaluator (GainRatioAttributeEval) and an attribute ranking algorithm (Ranker) in Weka [22], to characterize the most significant features that can be used for fingerprints. Gain-ratio rectifies the biasness of a feature with a maximum number of unique values and identifies the correlation of the features [23]. Features which are deemed insignificant, were subsequently removed from the selected features list based on the following criteria: gain ratio value <=0, multi-valued attributes (e.g. tcp.option_kind), time-dependent attributes (e.g. tcp.options.timestamp.tsval), attributes with negative/hexadecimal/binary values (e.g. udp.checksum), to select the most prominent subset of features. Empirically, 9 device-specific features, as listed in Table 2, were selected from the network and transport layers protocols to generate device fingerprints. For instance, some samples of data are presented in Table 2. These 9 features carry a device significant characteristics; for instance, resource-constraint IoT devices use smaller buffer size (tcp.window_size–it regulates a sender device is allowed to transmit data at a specific time, which impacts communication packets length) as compared to the general-purpose non-IoT devices [24], and IoT devices use a limited number of unique port numbers for their communication, whilst these devices communicate with a specific set of servers as compared with non-IoT devices [5], [16].

Table 2. Selected subset of network traffic 9 features with sample values (IoT Sentinel dataset)

| Tcp.srcport | Tcp.stream | Tcp.ack | Tcp.window_size | Udp.srcport | Udp.stream | Ip.len | Ip.ttl | Ip.proto | Class |
|---|---|---|---|---|---|---|---|---|---|
| 62997 | 0 | 0 | 8688 | | | 60 | 64 | 6 | Aria |
| 38067 | 56 | 4352 | 14048 | | | 366 | 64 | 6 | D-LinkCam |
| 38067 | 56 | 4352 | 14048 | | | 366 | 64 | 6 | D-LinkCam |
| | | | | 47581 | 653 | 65 | 64 | 17 | HueBridge |
| | | | | 47581 | 653 | 65 | 64 | 17 | HueBridge |
| | | | | 3074 | 1 | 443 | 4 | 17 | WeMoSwitch |
| 52266 | 22 | | | | | 76 | 64 | 17 | HueSwitch |

### 3.3. Machine learning classifiers

The selected features set was used for training six supervised ML classifiers: J48, RF, RT, VT, BG, and NB, to evaluate the proposed DFP scheme performance. These classifiers are described briefly, RF classifier is a supervised ML approach, which uses an ensemble of decision trees for decision-making [22], [25] and is capable of avoiding the over-fitting problem. The method can increase its accuracy by using majority voting for classification. RF classifier has been used in different research fields for various purposes, including devices classification [5], [16], and intrusion detection system (IDS) [26]. J48: The C4.5 algorithm [23] is an extended version of the iterative dichotomiser-3 (ID3) algorithm. In Weka, it is implemented as a J48 classifier [22]. The classifier produces a decision tree for classification based on information theory, and it has been utilized for various purposes, including devices classification [11], [20] and IDS [27]. RT classifier is a supervised ML algorithm, consisting of RF algorithms and a combination of single model trees, which perform non-pruning operations [22]. A large set of uniformly distributed RT assists in improving the model accuracy and avoid the over-fitting problem. Ensemble random trees, compute a final result (class-level) based on a majority voting mechanism for classification [28]. Vote classifier provides a way of combining multiple classifiers, such as J48 and BG, for classification [17], and regression approaches, by estimating decisions based on either average probability distributions or numeric projections of multiple classifiers [22]. BG is an ensemble of meta-algorithms, which is utilized for improving the model stability and accuracy. It is able to avoid overfitting problem as well as reduce variance. Instead of voting, the BG classifier predicts actual value based on average probability estimations for classification [22]. NB classifier is a probabilistic classifier based on the bayes theorem [29]. It uses normal distribution of all numeric attributes and kernel density estimators to improve the model performance [22], whereby the classifier computes each attribute value independently. It has been utilized for device identification [30], and anomaly detection.

### 3.4. Performance evaluation metrics

The significance of the empirically selected subset of features was evaluated based on different performance measures, including true positive rate (*TPR*) or recall, accuracy (*ACC*), and precision (*PRE*). These performance measures are defined,

$$TPR = \frac{TP}{TP+FN} \tag{1}$$

$$PRE = \frac{TP}{TP+FP} \tag{2}$$

$$ACC = \frac{(TP+TN)}{(TP+FP+TN+FN)} \tag{3}$$

where true-positive (*TP*) and true-negative (*TN*) are the total number of correctly classified positive and negative instances, respectively, according to predefined classes. On the other hand, false-positive (*FP*) and false-negative (*FN*) are the total amount of instances incorrectly predicted as positive and negative, respectively.

## 4. RESULTS AND DISCUSSION

The proposed IoT and non-IoT DFP model has been evaluated using the benchmark Weka tool [22]. Two online and an experimental dataset have been used to evaluate the performance of the proposed DFP scheme, based on the 6 ML classifiers. Device-specific features in Table 2 were extracted from the traffic traces, and used to train and test different classifiers. The dataset instances were randomly split into two subsets: 80% for training, and the remaining 20% for testing the DFP model.

### 4.1. Comparative classification results of the features

Using the proposed DFP model, a total of 9 features, as presented in Table 2, were extracted from both network and transport layers of the traffic traces. The significance of the features from the different layers have been analyzed, with Table 3 showing the performances of the proposed DFP model using the J48 classifier on different datasets, with features extracted from either the network layer only, transport layer only or combinations thereof. The J48 classifier was used, as it is the top performing classifier from the top 10 ML algorithms given in [31]. It can be seen that using only features from the transport layer (ACC 95.884%, TPR 82.23%) gives better performance as compared to using features from the network layer only (ACC 93.637%, TPR 39.50%) on the UNSW non-IoT dataset. Using all 9 features give the best performance, reaping the benefits derived from the unique characteristics from both layers, giving ACC of 99.692%, and TPR of 94.00%. Conversely, using 82 features from both layers the proposed DFP model reaches up to 98.23% accuracy, which is slight lower than the combined 9 features accuracy. The significance of the selected features subset of the proposed DFP model has also been evaluated using the J48 classifier on the IoT Sentinel dataset [14]. Similar to results on the UNSW non-IoT dataset, combining features from both network and transport layers gives the best results, with ACC of 91.143%, and TPR of 91.10%, whilst only 79.15% accuracy obtains using 82 features. The UNSW non-IoT and IoT Sentinel datasets have also been combined, and used to test the proposed DFP scheme, with similar results observed. The DFP model gains comparatively 1.69% lower accuracy using 82 features. Using features from the transport layer only gives better performance than using features from the network layer only, with combining features from both layers giving the best outcome. It can be concluded that combining all the 9 features in the proposed DFP model from both layers, extracts complementary information and consequently, gives the optimum performances, whilst insignificant features set reduce overall classification performances.

Table 3. Performance of the proposed DFP model in identifying with different set of features

| Dataset | Measures | Network layer (IP) 3 features | Transport layer (TCP-UDP) 6 features | Combined (TCP-UDP-IP) 9 features |
|---|---|---|---|---|
| UNSW non-IoT | ACC | 93.637% | 95.884% | 99.692% |
|  | TPR | 39.50% | 82.23% | 94.00% |
| IoT Sentinel | ACC | 49.779% | 63.406% | 91.143% |
|  | TPR | 49.80% | 63.40% | 91.10% |
| Combined UNSW non-IoT | ACC | 46.706% | 65.990% | 93.216% |
| and IoT Sentinel | TPR | 46.70% | 66.00% | 93.20% |

### 4.2. Device identification

The proposed DFP scheme has been assessed in terms of its ability to (i) distinguish device type, (ii) identifying individual IoT and non-IoT devices, using six different ML classifiers: J48, RF, RT, VT, BG, and NB. These are given in Figures 2 and 3, respectively. On the IoT Sentinel and UNSW non-IoT datasets as shown in Figure 2(a), it can be observed that the RF classifier gives the best performance, with average 99.3% and 100% precisions in identifying IoT and non-IoT devices, respectively. The lowest classification performance is obtained using the NB classifier, with an average precision of 58.3% for IoT devices and 98.4% for non-IoT devices. The proposed DFP method has also been evaluated using the UNSW dataset as shown in Figure 2(b). Again, the RF classifier gives maximum precision of 99.8% in identifying IoT and non-IoT devices, with the NB classifier giving the lowest performance. Other classifiers also give excellent results with over 99.5% and 99.9% precisions in identifying IoT and non-IoT devices, respectively. Figure 2(c) depicts performance on the IoT Sentinel and lab non-IoT datasets. Similar to previous observations, the RF classifier gives maximum precisions of 99.6% and 99.9% for IoT and non-IoT devices identification, respectively. Again, the worst performance is given by the NB classifier with a minimum of 96.5% precision for the identifying non-IoT devices and 74.2% precision for IoT devices. Generally, other classifiers also show good performances.

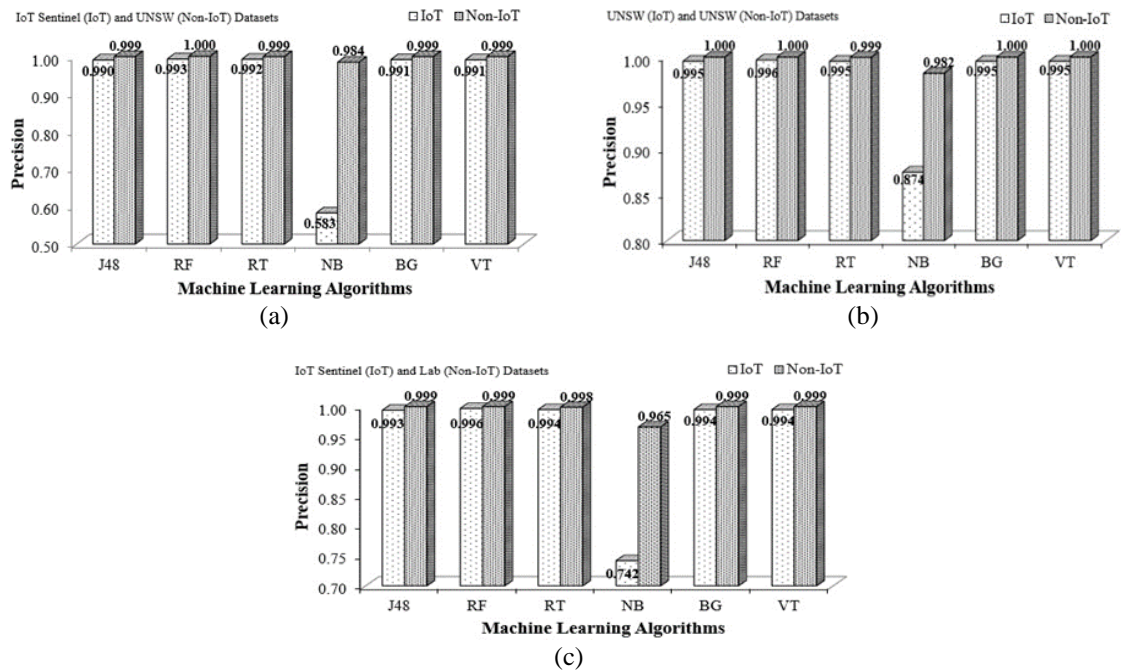

Figure 2. Device type classification performance on: (a) IoT Sentinel and UNSW non-IoT datasets, (b) UNSW dataset, and (c) IoT Sentinel and lab datasets

The performances of the proposed DFP model in identifying individual devices on different datasets containing non-IoT devices, and combination of IoT and non-IoT devices using various ML algorithms, are presented in Figure 3. On the UNSW non-IoT dataset as shown in Figure 3(a) consisting of 7 non-IoT devices, the RF classifier gives the highest average precision of 98.2%, whilst the NB classifier gives the lowest average precision of 44.7%. Other classifiers give respectable performance with over 94.4% precision. On the Lab dataset as shown in Figure 3(b), all classifiers give over 98.8% average precision, except for the NB classifier, which only manages 77% precision. The RF classifier gives the highest precision of 99%. Again, the NB classifier again gives the lowest precision performance of only 31.5% on the combined IoT Sentinel and UNSW non-IoT datasets in Figure 3(c), with the RF, BG, and VT classifiers giving up to 91.6% average precisions. Finally, similar observation is observed on the combined datasets from IoT Sentinel and lab datasets in Figure 3(d), with the RF classifier gives the highest precision of 95.7%, and other algorithms giving respectable precision values of over 90.2%, except the NB classifier. The NB classifier only manages 43.5% precision.

### 4.3. Comparative analysis of the proposed model

Table 4 shows a comparison of different DFP models. The proposed scheme achieves good performance, by extracting only 9 selected features to serve as device-specific signatures, from a single TCP/IP packet information. This is in contrast to the method proposed by Aksoy and Gunes [20], whilst the researchers utilized a total of 212 features from a packet. Similarly, in reference [11], the authors generate fingerprints using 161 features from a single packet. Sivanathan *et al.* [5], [16], which requires statistical analysis to generate device signatures from *N* number of packets, depending on the traffic flow. Researchers [6], [16]–[19] have focused on distinguishing device type, and identifying individual IoT devices only. On the other hand, the proposed DFP method assists not only in distinguishing device type, but also allows the identification of both IoT and non-IoT devices with high accuracy, by utilizing a shortlist of features set from packet information. The proposed DFP model uses real-values of the individual features, to generate device-specific signatures, whilst prior works have either utilized statistical features or a combination of statistical and real-value related features for device fingerprints, which consequently, require a longer time to generate.

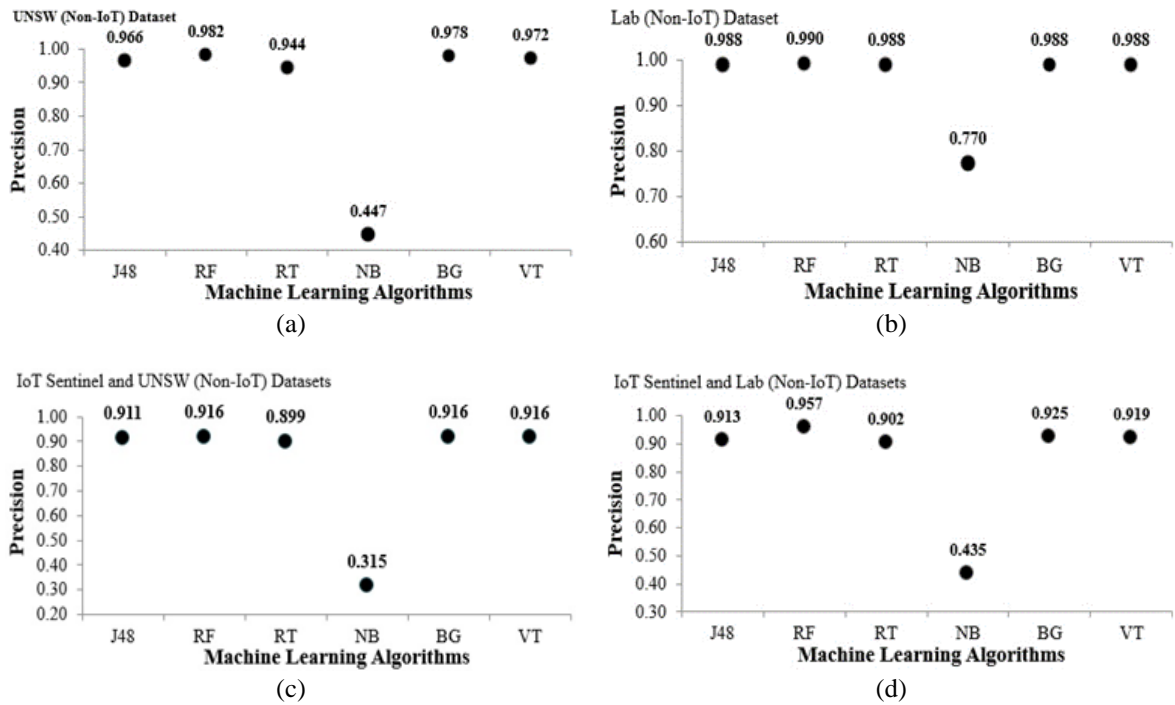

Figure 3. IoT and non-IoT devices classification performances on: (a) UNSW non-IoT, (b) Lab non-IoT, (c) IoT Sentinel and UNSW non-IoT, and (d) IoT Sentinel and lab datasets

Table 4. Comparison of the device fingerprinting models

| Source | Category | | Individual device | | Packet | Features | Feature analysis | Datasets | |
|---|---|---|---|---|---|---|---|---|---|
| | IoT | Non-IoT | IoT | Non-IoT | | | | | |
| Sivanathan *et al.* [16] | ● | ● | | | | *N* (Flow) | 12-Stat. | Traffic analysis (Communication patterns) | O, P |
| Sivanathan *et al.* [5] | ● | ● | | | | *N* (Flow) | 8-Stat. | | O, P |
| Bremler-Barr *et al.* [6] | ● | | | | | *N* (20-Min.) | 10-Stat. | F1-score > 0.5 | O, P |
| Pinheiro *et al.* [17] | ● | ● | | | | *N* (1-Sec.) | 3-Stat. | -- | P |
| Ortiz *et al.* [18] | ● | ● | | | | *N* (Flow) | Auto | Auto-encoder | P |
| Meidan *et al.* [19] | ● | ● | | | | *N* (Flow) | 3-Stat. | -- | O, P |
| Aksoy and Gunes [20] | | | ● | | | 1 | *212-Val. | *One-level (None) Two-level (genetic algorithm) | P |
| Chowdhury *et al.* [11] | | | ● | | | 1 | 161-Val. | Metric entropy | P |
| This work | ● | ● | ● | | | 1 | 9-Val. | Gain ratio, ranker | O, P |

## 5. CONCLUSION

The number of both IoT and non-IoT devices being deployed in the cyberspace to connect to various services via the internet, is increasing rapidly, due to the reduced costs of hardware and improved affordability of internet access. It is becoming challenging to distinguish between device type, and even more challenging to identify individual devices using user-defined explicit identifiers, such as IP/MAC addresses. In this study, an efficient DFP model has been proposed based on packets header protocols information from both the network and transport layers. Empirically, 9 device-specific features from a single TCP/IP packet of information have been selected to be used as device fingerprints. The selected features are not dependent on user-defined identifiers, and are note easily mutable. Hence, the features can be used to increase the level of security in a network, and can help in mitigating different security threats, including spoofing attacks. These features can be used to distinguish device type, as well as to identify individual IoT and non-IoT devices. Experimental results have shown that combining features from both the network and transport layers provides the best results. Additionally, it has been shown that RF classifier gives the highest performance, whilst NB classifier gives the worst performance among all the considered classifiers. The proposed DFP method is able to distinguish device type with up to 100% precision, as well as able to identify individual devices with up to 95.7% precision. The results are significant, as the proposed DFP model can be implemented to improve the security of networks.

## BIOGRAPHIES OF AUTHORS

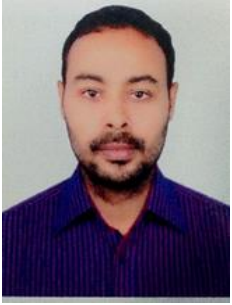

**Rajarshi Roy Chowdhury** is currently pursuing his PhD in Systems Engineering under the Faculty of Integrated Technologies (FIT), Universiti Brunei Darussalam (UBD). He obtained his Master's degree in Computer Science from Universiti Sains Malaysia (USM), Malaysia in 2012. Later, he joined Sylhet International University (SIU), Bangladesh, as a Lecturer in 2012. His research interests are internet of things (IoT), networking, and machine learning (ML). He can be contacted at email: rajarshiry@gmail.com or 19h0901@ubd.edu.bn.

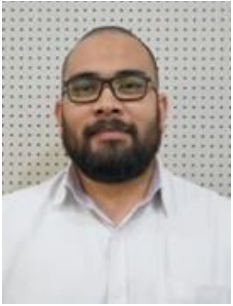

**Dr. Azam Che Idris** is a chartered engineer with a wide interest in technology. Originally trained in high-speed aerodynamics, he gained major exposure to IR4.0 technology during his tenure in a defence consultancy group. His current interest is utilizing machine learning (ML) to understand hypersonic flow physics and to control air-breathing engine in Mach 5. He holds a doctorate in Aerospace Engineering from University of Manchester, UK. He can be contacted at email: azam.idris@ubd.edu.bn.

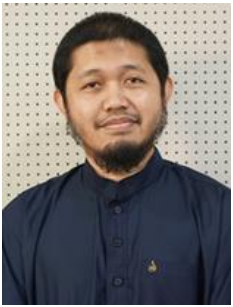

**Dr. Pg Emeroylariffion Abas** received his B.Eng. Information Systems Engineering from Imperial College, London in 2001, before obtaining his PhD Communication Systems in 2005 from the same institution. He is now working as an Assistant Professor in System Engineering, Faculty of Integrated Technologies, Universiti Brunei Darussalam. His present research interest are data analysis, security of info-communication systems and design of photonic crystal fiber in fiber optics communication. He can be contacted at email: emeroylariffion.abas@ubd.edu.bn.